\documentclass[aps,prx,twocolumn,showpacs,floatfix,nobibnotes,nofootinbib,superscriptaddress,amsmath,amssymb,longbibliography]{revtex4-1}
\usepackage[english]{babel}
\usepackage[colorlinks,urlcolor={blue}]{hyperref}
\usepackage{eurosym}
\usepackage[usenames]{color}
\usepackage{graphicx}
\usepackage{amsmath}
\usepackage{amsfonts}
\usepackage{amssymb}
\usepackage{mathrsfs}
\usepackage{bm}
\usepackage{verbatim}
\usepackage{ulem}

\usepackage[textsize=small]{todonotes}

\newcommand{\beq}{\begin{equation}}
\newcommand{\eeq}{\end{equation}}
\newcommand{\bea}{\begin{eqnarray}}
\newcommand{\eea}{\end{eqnarray}}

\begin{document}

\title{A further study on the renormalization group aspect of perturbative corrections}
\author{C.-J.~Yang}
\affiliation{ELI-NP, ``Horia Hulubei" National Institute for Physics and Nuclear Engineering, 30 Reactorului Street, RO-077125, Bucharest-Magurele, Romania}
\email{chieh.jen@eli-np.ro}

\begin{abstract}
I perform a further study regarding a renormalization-group (RG) issue---which concerns a wide variety of the so-called perturbative power counting under effective field theories (EFT)---as pointed out by A. M. Gasparyan and E. Epelbaum [Phys. Rev. C 107, 034001 (2023)]. I show that the issue could originate from a wrong power counting, or from treating those incomplete, truncated amplitudes beyond the degree to which they should be trusted. 
Meanwhile, under EFT principles, one should
always associate the result with an uncertainty that is adequate to its EFT order. One way to
accommodate this is to encode its effect in a more general form of contact terms. In this regard, no
RG issue is found in the $^3$P$_0$ nucleon-nucleon scattering under the Long and Yang power counting. In contrast, the RG issue under Weinberg's pragmatic proposal remains a problem even with uncertainty taken into account.

\end{abstract}

\maketitle

\section{Introduction}
One of the most important developments in modern theoretical physics is the conception of effective field theory (EFT). Pioneered by Weinberg~\cite{Weinberg:1978kz,Coleman1969,Callan1969b,Weinberg1999}, the key concept is to arrange physical observables order-by-order based on an appropriate effective Lagrangian that accommodates the energy scale of interest and the symmetries of the system that one wishes to describe. In contrast to an exact theory---where everything needs to be specified or known---EFT accommodates practical difficulties such as the impossibility of carrying out all of the higher-order calculations, the ignorance of the unknown physics, or both. This leads to the philosophy that physical descriptions should be improved order-by-order based on available information~\cite{Hartmann2001}. The ignorance of higher-order corrections (regardless of their origin) is compensated by renormalization and is accompanied by the entrance of low-energy constants (LECs). LECs are free parameters in the theory and are normally fitted to observables in order to obtain reasonable descriptions. However, apart from a few cases where physical observables can be obtained analytically such as nuclear forces based on pionless EFT up to the few-body level~\cite{Hammer:2019poc} or certain quantum electrodynamics (QED) processes, the fitting processes interfere with the prediction, which can potentially grant inconsistent theories and prevent us from a deeper understanding of nature\footnote{This is also subjected to viewpoint of what is ``ab-initio"~\cite{Machleidt:2023jws,Ekstrom:2022yea}.}. Therefore, when numerical solutions are the only possibility, a careful check of the renormalizability of the final results is required. Only after that could one justify whether a consistent approach based on the principles of EFT is obtained. 

In nuclear EFT, it is known that currently the most popular arrangement of nuclear forces---Weinberg’s pragmatic proposal (WPP)~\cite{We90,We91}---fails to generate renormalizable results~\cite{Hammer:2019poc,Griehammer2022}. This is demonstrated by the breakdown of the renormalization group (RG) invariance of the nucleon-nucleon (NN) scattering amplitude from the leading order (LO) throughout next-to-next-to-next-to-leading order (N$^3$LO)~\cite{Nogga:2005hy,PhysRevC.74.054001,PhysRevC.74.064004,Yang:2009kx,Yang:2009pn,Zeoli2013-ro}, and up to next-to-leading order (NLO) in few-body systems~\cite{Yang:2020pgi,Yang:2021vxa}.

As a consequence, studies~\cite{ksw,ksw1,Birse:2007sx,Long:2007vp,Valdper,Valdperb,BY,BYb,BYc,bingwei18,Song:2016ale,Peng:2020nyz,Habashi:2021pbe,Yang:2020pgi,Yang:2021vxa,Thim:2023fnl,Li:2023hwe,Thim:2024yks,Thim:2024jdv} suggest that a consistent
treatment of nuclear forces involves a non-perturbative treatment only at LO.
Higher-order corrections are to be added perturbatively
under distorted-wave-Born-approximation (DWBA).
Note that this treatment has been applied in the context of pionless EFT, where, in addition to numerical analysis, analytic or semi-analytic expressions show strong evidence that RG-invariance is satisfied~\cite{vanKolck:2020llt,vanKolck:2020plz}. A recent Bayesian analysis also indicates that its breakdown scale is consistent with $M_{hi}\approx m_{\pi}$~\cite{Ekstrom:2024dqr}.

Despite the above-demonstrated success of the DWBA-based power counting (PC), there are controversies~\cite{de5,de6,de7,de8}. On one hand, numerical analyses show that the RG issues of WPP become severe once the ultraviolet cutoff $\Lambda$ associated with the non-perturbative iteration is increased beyond $1$ GeV. This leads Refs.~\cite{de1,de2,de3,de4} to argue that one should stay within a limited cutoff window (which normally ranges from $450 \sim 600$ MeV), as the non-perturbative treatment at LO already mandated the “peratization” of an EFT for an arbitrarily high cutoff. On the other hand, results generated by PC based on DWBA have certain features that can be either inconvenient or require further studies. The inconvenient feature concerns the possible need to restore missing pole positions (mainly the bound-states) under perturbation theory~\cite{pionless16,pionless16b,Yang:2020pgi,Yang:2023guo,Contessi:2023yoz,Contessi:2024vae}. Moreover, a recent study by Gasparyan and Epelbaum~\cite{Gasparyan:2022isg} discovered an interesting feature of perturbative treatment, which leads the authors to claim that a wide variety of DWBA-based PCs are generally not RG invariant. The authors demonstrated their argument by explicit calculations on the NN scattering problem, under a toy model as well as the PC of Long and Yang~\cite{BY,BYb,BYc}. In response, Ref.~\cite{Peng:2024aiz} advocates that the issue can be avoided by adjusting the fitting strategy to make those exceptional cutoffs amenable to chiral effective field theory.

In this work, the aforementioned RG issue regarding perturbative PCs is further studied under the EFT principles. In particular, I will show that the analysis performed in Ref.~\cite{Gasparyan:2022isg} does provide a new and valuable methodology in the aspect of PC-analysis. Nevertheless, it needs to be accompanied by the ingredient that any EFT calculation should be associated with an uncertainty adequate to the given order. Once the uncertainty prescribed by the PC is taken into account, the appearing RG issue disappears in the $^3$P$_0$ case for the Long and Yang PC~\cite{BY}.

The paper is organized as follows. In Sec.~\ref{sec1}, I argue conceptually that the RG-analysis performed in Ref.~\cite{Gasparyan:2022isg} could suffer from the problems of treating the incomplete, truncated amplitudes exactly. A refinement for it to become a meaningful tool in analyzing PCs under EFT is proposed. 
Then I demonstrated the concept with a toy model in Sect.~\ref{1s0}. In Sec.~\ref{3p0}, the PC of Long and Yang in the NN $^3$P$_0$ channel is analyzed using a more general form of contact terms. In Sec.~\ref{wpc}, the analysis is applied to WPP. 
Finally, the main point is stressed in Sec.~\ref{regu} and the findings are summarized in Sec.~\ref{con}.

\section{RG-analysis with uncertainty taken into account}
\label{sec1}

\subsection{General consideration}
An EFT-justified calculation must have results organized in an order by order improvable manner. Denoting $\mathcal{O}_n$ the observable evaluated up to order $n$, it must scale as\cite{Griesshammer:2015osb,Griesshammer:2020fwr,yang_rev}: 
\begin{align}
&\mathcal{O}_n(M_{lo};\Lambda;M_{hi})=\underbrace{\sum_i^n\left(\frac{M_{lo}}{M_{hi}}\right)^i\mathcal{F}_i(M_{lo};M_{hi})}_{\text{trustable part}}\notag \\ 
&+\underbrace{\operatorname{\mathscr C}_n(\Lambda;M_{lo},M_{hi})\left(\frac{M_{lo}}{M_{hi}}\right)^{n+1}}_{\text{uncertainty}},
\label{pc}
\end{align}
where $M_{lo}$ denotes the low-energy scales, and $M_{hi}$ is the breakdown scale. The ``trustable part" associated with $\mathcal{F}_i$ denotes processes that have been explicitly evaluated. The evaluation includes renormalization/fitting to observables. 
Meanwhile, as long as one stops at a finite $n$, results will always contain a residue $\operatorname{\mathscr C}_n$, which accounts for the ignorance of higher-order effects. Although the exact form is unknown, for the expansion to work, the uncertainty must contain a suppression $\sim\left(\frac{M_{lo}}{M_{hi}}\right)^{n+1}$, which means that one is more sure about the answer $\mathcal{O}_n$ as the order $n$ increases. $\operatorname{\mathscr C}_n$ must not scale as positive powers of $\Lambda$, otherwise RG is broken. Normally, $\operatorname{\mathscr C}_n$ oscillates with $\Lambda$ and converges to a constant such that $\left(\frac{\mathscr{C}_n}{\mathcal{F}_n}\right)\sim O(1)$ after $\Lambda \gg M_{hi}$. In other words, ``weak naturalness" is required in order for the series to converge~\cite{Wesolowski2016,Melendez:2020ikd}. Since one usually demands $\mathcal{O}_n$ to fall in the vicinity of experimental data $\mathcal{O}_{exp}$, the LECs are adjusted in a way that the difference between the ``trustable part" and $\mathcal{O}_{exp}$ is remediable by the ``uncertainty". After adjustment, the effects due to different fitting strategies, $\Lambda$, or the regulator on $\mathcal{O}_n$ should be equal to or smaller than the ``uncertainty" part. This also means that a successful EFT only concerns/ensures that $\mathcal{O}_n$ is trustable up to an uncertainty that is adequate to the considered order, rather than having an exact value.

\subsection{Features at LO with the presence of a singular attractive potential}
\label{lo1}

At LO, a non-perturbative treatment is necessary for at least part of the interactions. This non-perturbative treatment also gives rise to bound-states. Denoting $V_{LO}$ as the LO interaction, the LO amplitude can be obtained by solving the Lippmann-Schwinger equation (LSE), i.e.,
\begin{widetext}
\begin{equation}
T_{LO}(p^{\prime },p;E)=V_{LO}(p^{\prime
},p)+\frac{2}{\pi }M\int_{0}^{\infty }\frac{
dp^{\prime \prime }\;p^{\prime \prime }{}^{2}\;V_{LO}(p^{\prime },p^{\prime \prime })\;T_{LO}(p^{\prime \prime
},p,E)}{p_{0}^{2}+i\varepsilon -p^{\prime \prime }{}^{2}},
\label{eq:2.3}
\end{equation}
\end{widetext}
where $p_{0}^{2}/M=E_{cm}$ is the cm energy, $M=939$ MeV is the nucleon mass.

Adopting the short-hand operator form as described in Ref.~\cite{BY}, the NLO correction is 
\begin{align}
T_{NLO}& =V_{NLO}+V_{NLO}GT_{LO}+T_{LO}GV_{NLO} \nonumber\\
& + T_{LO}G\,V_{NLO}\,GT_{LO}\, ,   \label{eqn:LSE23}
\end{align}
with
\begin{equation}
G=\frac{2}{\pi }M\int_{0}^{\infty }\frac{ p^{\prime \prime }{}^2dp^{\prime
\prime }}{p_{0}^{2}+i\varepsilon -p^{\prime \prime }{}^{2}}.
\end{equation}
One can denote the LO intereaction as
\begin{equation}
V_{LO}=[V^{(L)}_{LO}+V^{(S)}_{LO}]f_R(\Lambda), 
\label{vlo} 
\end{equation}
where $f_R(\Lambda)$ is a regulator of choice, and $V^{(L)}_{LO}$ and $V^{(S)}_{LO}$ are the long- and short-range components, respectively. Note that the operator form in the above equations is adopted, with the $(p',p)$ applied implicitly.

For convenience, one can further define 
\begin{equation}
V^{(S)}_{LO}=\sum_{i=1}^{d_{lo}} C_{i}v_i, 
\label{vs} 
\end{equation}
where $C_{i}$ is the LEC, $d_{lo}$ is the upper limit of the derivative at LO, and $v_i$ is the corresponding operator structure.
For example, with $V^{(L)}_{LO}$ being the one-pion-exchange potential (OPE), $V^{(S)}_{LO}=C_T$ in $^3$S$_1$-$^3$D$_1$ and $V^{(S)}_{LO}=C_Ppp'$ in $^3$P$_0$ channel. 

Note that the NN potential can be singular and attractive in certain partial-waves, and it is known that the LEC within $V^{(S)}_{LO}$---which is adjusted to renormalize $V^{(L)}_{LO}$---then has a limit-cycle behavior~\cite{Long:2007vp} and diverges at specific values of $\Lambda$. This is illustrated below in Fig.~\ref{fig1} and Fig.~\ref{fig2}. 

\begin{figure}[h]
\includegraphics[width=0.47\textwidth,clip=true]{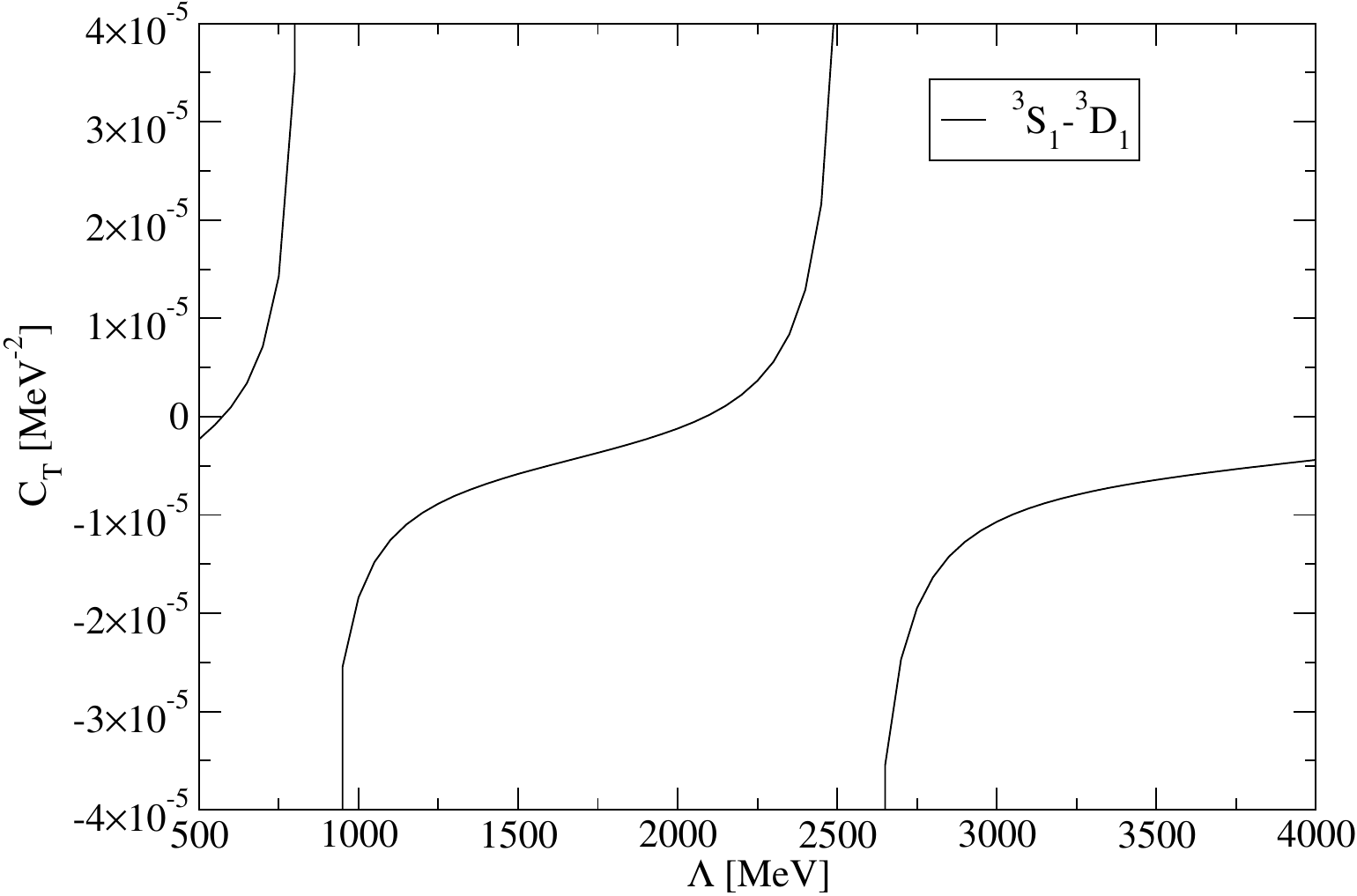}
\caption{The renormalization constant $C_T$ for the LO $^3$S$_1$-$^3$D$_1$ channel as a
function of cutoff $\Lambda$, where a sharp cutoff $f_R(\Lambda)=\theta(\Lambda)$ is adopted. This plot is taken from Fig.~2 of Ref.~\cite{Yang:2007hb}. }
\label{fig1}
\end{figure}

\begin{figure}[h]
\includegraphics[width=0.5\textwidth,clip=true]{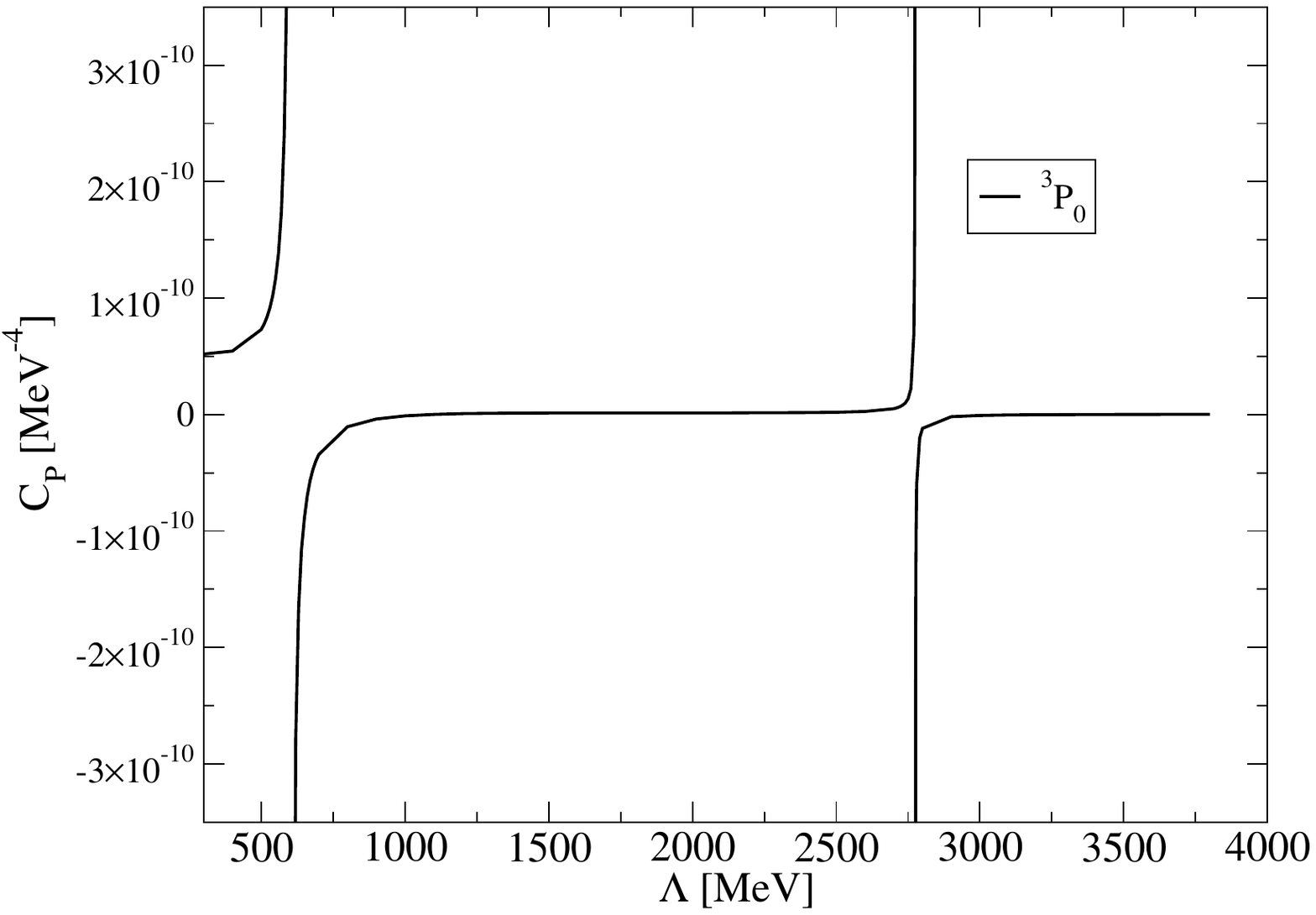}
\caption{The renormalization constant $C_P$  for the LO $^3$P$_0$ channel  as a
function of cutoff $\Lambda$, where a sharp cutoff $f_R(\Lambda)=\theta(\Lambda)$ is adopted.  }
\label{fig2}
\end{figure}

Note that the non-perturbative treatment is equivalent to an eigen-value problem with Hamiltonian $H=KE+V_{LO}$, where $KE$ denotes the kinetic term. After diagonalization, one obtains the LO wavefunction $|\psi_{LO}\rangle_i$ and eigenvalues $E_{i}=p_{i}^2/M$. $E_i$ is continuous for the scattering states.
At cutoffs ($\equiv\Lambda^*$) where the LEC ($\equiv C_x$) diverges (e.g., $\Lambda\approx 700, 2600$ MeV in Fig.~1), $|\psi_{LO}\rangle_i$ has a special property, i.e.,
\begin{align}
&\langle \psi_{LO}|v_xf_R(\Lambda^*)|\psi_{LO}\rangle_i=0, \text{ for all }p_i, \nonumber\\
C_x &\langle \psi_{LO}|v_xf_R(\Lambda^*)|\psi_{LO}\rangle_i=\text{finite number, for all } p_i .
\label{zero} 
\end{align}
The peculiar feature of Eq.~(\ref{zero}) is guaranteed since all eigenvalues $E_i$---which correspond to $\langle H\rangle_i$---are finite. Then $\langle KE\rangle_i$ is obviously finite and $\langle H\rangle_i= \langle KE\rangle_i+\langle V_{LO}\rangle_i$. Thus, $\langle V_{LO}\rangle_i$---which is the term on the left side in the second line of Eq.~(\ref{zero})---must be finite also.  

It needs to be stressed that the key, i.e., the non-perturbative treatment, must be employed for the above to hold. 

However, if a new interaction $V_{new}$ is adopted---which can contain a set of operators with a structure different from $V^{(S)}_{LO}$ (e.g., higher derivative terms), or with the same structure of $V^{(S)}_{LO}$ but with different regulators or cutoff (e.g., $\Lambda_{NLO}\neq \Lambda_{LO}$)---then its matrix element is no longer protected by the eigenvalue properties. 
This is the very origin of the RG issue for DWBA-based PCs as investigated in Ref.~\cite{Gasparyan:2022isg}, and will be analyzed in further detail in the following sections.

\subsection{Dilemma at next-to-leading order for DWBA} 
Consider now the next-to-leading order (NLO) correction from DBWA. The calculation follows perturbation theory and involves the matrix elements consisting of NLO operators sandwiched by $|\psi_{LO}\rangle_i$. Define the NLO interactions as
\begin{equation}
V_{NLO}=[V^{(L)}_{NLO}+V^{(S)}_{NLO}]f_R(\Lambda),
\label{vnlo} 
\end{equation}
where
\begin{equation}
V^{(S)}_{NLO}=\sum_{i=1}^{d_{nlo}} C^{(nlo)}_{i}v_i, 
\label{vnlos} 
\end{equation}
with $C_i^{(nlo)}$ the NLO LECs, and $d_{nlo}\geq d_{lo}$ the upper limit of the derivative at NLO. 
The NLO correction to $E_i$ is then 
\begin{equation}
E_i^{NLO}=\langle\psi_{LO}|V_{NLO}|\psi_{LO}\rangle_i. \label{enlo} 
\end{equation}
In general, the short-range part of the NLO interaction, $V^S_{NLO}$, has a richer structure than its LO correspondence. Thus, the special property as listed in Eq.~(\ref{zero}) will no longer hold.  Consequently, the matrix element of a new NLO operator can be zero only at $p_0=p_x$, but finite elsewhere. This leads to a dilemma if one chooses to renormalize the LEC to an observable at $p_x$. On one hand, for this term to have a non-zero correction at $p_x$, the corresponding LEC would need to be $\infty$. However, since Eq.~(\ref{zero}) does not apply anymore, a diverged LEC will cause the results to diverge for all other $p_0\neq p_x$. More generally, when a set of operators enters (which is standard for NLO and higher orders), their linear combination, when coupled with the ingredient that at least a subset of the LECs are fixed by a particular renormalization procedure, leads to the same dilemma. The above is referred to as ``not factorizable zero" in Ref.~\cite{Gasparyan:2022isg}. 

As long as $V_{LO}$ is singular and attractive, the LO wavefunction will oscillate more and more at a shorter distance (or equivalently, higher momentum) so that in general one can always find a $p_x$ where the ``not factorizable zero" occurs at NLO.
In practice, the cutoff window where $p_x$ falls into the physical region (below $M_{hi}$) can be rather narrow. Nevertheless, it exists. It is demonstrated in Fig.~5 of Ref.~\cite{Gasparyan:2022isg} that the ``exceptional cutoff" ($\equiv\Lambda_e$) occurs between 2.6990 and 2.6991 GeV for the PC of Long and Yang at NLO. 

\subsection{Origin of the dilemma and its solution}
\label{2d}
Ref.~\cite{Gasparyan:2022isg} considered the above dilemma as an RG problem, arguing that the origin of it is due to that the non-perturbative LO amplitude is not ``properly renormalized". In their following works~\cite{PhysRevC.105.024001,Gasparyan:2024qbi}, the authors further explicated that the non-perturbative treatment at LO imposes ``nontrivial constraints on a choice of the effective interaction and the renormalization scheme", and one would need to perform subtractions in all possible subdiagrams in the spirit of the
Bogoliubov-Parasiuk-Hepp-Zimmermann (BPHZ) renormalization procedure to ensure renormalizability of subleading results.

However, one should keep in mind that any fitting procedure of LECs should always take uncertainty into account.
Moreover, within DWBA, the NLO correction is evaluated by sandwiching the NLO operators with the LO wavefunction $|\psi_{LO}\rangle$---which itself comes with a theoretical uncertainty of size $\lesssim |\psi_{NLO}\rangle$. In fact, the starting point of DWBA already prescribes that all amplitudes are imperfect at each order. 
For example, when the S-matrix is not unitary (which could occur under non-relativistic DWBA or other perturbative treatments), one cannot force it to reproduce the data (e.g., phase shift $\delta$, cross section, etc.) exactly. Enforcing an exact fit to data would render the LECs to become complex. Thus, $T = \frac{e^{i\delta} \sin[\delta]}{M p_0}$ no longer holds order-by-order.
In this case, one has to anticipate an additional step to convert the amplitude/T-matrix perturbatively as well.
Explicitly, one expands 
\begin{align}
    T_{\text{up to O(n)}}=T^{(0)}+T^{(2)}+...+T^{(n)},\nonumber\\
    \delta_{\text{up to O(n)}}=\delta^{(0)}+\delta^{(2)}+...+\delta^{(n)}, \label{eq1}
\end{align}
in order to convert the T-matrix into phase shift and perform renormalization (see e.g., Eq.~(A1) in the appendix of~\cite{BY}).

Note that in the above expansion, not only the T-matrix is expanded and truncated, but the phase shift $\delta$ it supposed to reproduce is truncated also. 
Thus, the correspondence between LECs to observables is only approximately, as one can argue that fitting to any $\bar{\delta}= \delta_{\text{up to O(n)}}+(\text{a small term}\approx\delta^{(n+1)})$,
or equivalently, associating a small $\tilde{T}\approx T^{(n+1)}$ to the calculated T-matrix is also a valid renormalization. 
Without knowing the order ``$n$" in $T_{\text{up to O(n)}}$ of Eq.~(\ref{eq1}), the conversion to $\delta$ cannot be pinned down---a strike difference to the non-perturbative conversion. One could go on and specify the conversion detail in the above perturbative expansion and obtain an one-to-one correspondence between $T^{(n)}$ and $\delta^{(n)}$. But this one-to-one conversion is obtained by ignoring higher-order terms in Eq.~(\ref{eq1}) and is not exact.
Thus, although $T$ is improvable order-by-order, it is defective at any finite order (not just because of the ignorance of higher-order diagrams, but intrinsically in terms of failing some basic requirements of the scattering process). Consequently, adjusting LECs under DWBA up to NLO to reproduce a real value of $\delta(E=E^*)$ exactly is like asking $|T_{\text{up to O(2)}}=T^{(0)}+T^{(2)}|^2$ to fit $|T_{exp}|^2$, which is formally impossible even at a single energy $E=E^*$\footnote{$|T_{exp}|^2$ is directly related to the cross section $\sigma_{exp}$ via $|T_{exp}|^2=\frac{\sigma_{exp}}{4\pi(2l+1)M^2}$, with $l$ the angular momentum quantum number.}. In short, an uncertainty (as large as, but in addition to the truncation at diagram/loop level) appears explicitly due to the DWBA treatment, which should be taken into consideration when one wishes to perform further actions based on the resulting T-matrix. 

Note that one can compare the above situation to the case of a general fit, where it is also impossible to fit two data points exactly with one LEC (or data within a certain range with a finite set of LECs). In this case, one performs the ``best fit". And since the $\chi$-square value will be large should the phenomenon of ``exceptional zero" (including its nearby vicinity) occur, the corresponding LEC sets will automatically be excluded in the fitting process. 

Thus, although it is not commonly required (neither practiced in Ref.~\cite{BY,BYb,BYc}), formally, one should always include this uncertainty in the renormalization/fitting procedure, instead of directly solving a set of linear equations to obtain the LECs exactly. Adopting an exact procedure/fitting to perform the NLO (or higher-order) renormalization is equivalent to assuming an ``exact" wavefunction at the previous order---which violates the initial assumption of DWBA and EFT itself.

Back to the aforementioned issue of Long and Yang PC, the fact that one needs to fix the fitting procedure and fine-tune the cutoff to a precision more than five significant digits for the resulting LECs to give the problematic phase shift (as shown in Fig.~5 and Fig.~6 in Ref.~\cite{Gasparyan:2022isg}) is an indication that the RG issue might come from an illegal demand that requires DWBA to be more accurate than what it should be at the order considered\footnote{One should allow at least $10\%$ uncertainty for all NLO observables, judging from the estimate that $M_{lo}/M_{hi}\sim1/3$~\cite{PhysRevC.104.064001}.}. 
Note that such a fine-tune-related issue is different from the scenario of a PC failure. When a PC fails, the ``uncertainty" part in Eq.~(\ref{pc}) is either not under control at all (e.g., diverges / oscillates with $\Lambda$, such as WPP at LO in singular and attractive channels) or scales worse than the prescribed PC (which could be the case of the NN $^1$S$_0$ channel under WPP or the PC of Ref.~\cite{BYc}\footnote{Both are suggested by the slow converge pattern of the LO amplitude.}). On the other hand, a fine-tune-related problem has the ``uncertainty" part under control, and the problem only occurs when one demands an accuracy exceeding the theoretical uncertainty. In other words, even if the ``uncertainty" part of Eq.~(\ref{pc}) does indeed contain some undesirable properties due to a ``peratization" of the LO amplitude\footnote{The ``peratization"~\cite{de1}, if exist, must reside in the ``uncertainty" part, as the DWBA-based LO and NLO amplitudes are shown to satisfy RG- and PC-requirements by Refs.~\cite{Nogga:2005hy,Song:2016ale} and Sec. IV of the present work.}, as long as it is not exactly picked up and exploited to ruin the NLO renormalization, the EFT expansion can still be improved order by order.

One way to verify whether the RG problem is due to a PC failure or the illegal demand for precision is to take the uncertainty into account when performing the RG analysis.
This can be achieved by encoding the uncertainty into a more general form of regulator. 
The simplest case will be a combination of two regulators, i.e.,
\begin{equation}
\mathfrak{F}_R(\Lambda)\equiv xf_a(\Lambda)+(1-x)f_b(\Lambda), \label{r}
\end{equation}
where $f_a$ and $f_b$ are regulators with a slight difference. They should be chosen appropriately so that varying the parameter $x$ between 0 and 1 will induce an uncertainty on $\mathcal{O}_n$ (observables calculated up to order $n$) smaller or equal to the value prescribed by the PC. Note that $x$ depends on $\Lambda$, i.e., $x=x(\Lambda)$. However, it is not an LEC, but a means to incorporate uncertainty, as a concrete choice of its value should not be crucial. More generally, $x$ can be associated with a width and more than two regulators can be adopted. Varying $x$ within its width should result in a band in the observables which corresponds to the EFT uncertainty at the given order. Eq.~(\ref{r}) presents a simplified version that is easier to implement in numerical calculations. Thus, $x$ is merely a free and extra parameter that can be utilized to avoid the aforementioned fine-tuning related problems.  
Since $\mathfrak{F}_R(\Lambda)$ in Eq.~(\ref{r}) is linear in $x$, to check whether $f_a$ and $f_b$ are chosen properly for a given order $n$, it is sufficient to evaluate whether the difference between the two extreme cases is smaller than the prescribed uncertainty, that is,
\begin{widetext}
\begin{equation}
|\langle\psi_{LO}|(V^{L}_{NLO}+V^{S}_{NLO})\left(f_a(\Lambda)-f_b(\Lambda)\right)|\psi_{LO}\rangle_i|\lesssim \mathcal{O}_n(M_{lo};\Lambda;M_{hi})\left(\frac{M_{lo}}{M_{hi}}\right)^{n+1}, \text{ for all } p_i\leq M_{hi}.
\label{dr}
\end{equation}
\end{widetext}
Note that one caveat in the above equation is that $V^S_{NLO}$ contains LECs, and a diverged LEC could cause the matrix element to blow up, rendering Eq.~(\ref{dr}) useless. 
As the purpose of varying $x$ from 0 to 1 is to generate an uncertainty permitted at a given order in the matrix element, any check using Eq.~(\ref{dr}) should be evaluated at a typical cutoff where
the size of $V^L_{NLO}$ is comparable to $V^S_{NLO}$.


\section{Toy model in NN $^3$P$_0$ scattering}
\label{1s0}

I now incorporate Eq.~(\ref{dr}) to examine the toy model specified in Ref.~\cite{Gasparyan:2022isg}.
The LO potential is
\begin{equation}
V^{(^3P_0)}_{LO}=[V^{(^3P_0)}_{1\pi}+C_P pp']f_R(\Lambda_0),
\label{toy1}
\end{equation}
where $V^{(^3P_0)}_{1\pi}$ is the OPE in the $^3$P$_0$ channel.
At NLO,
\begin{equation}
V^{(^3P_0)}_{NLO}=[\bar{V}^{(^3P_0)}_{2\pi}+C^{NLO}_0pp']f_R(\Lambda_2),
\label{toy2}
\end{equation}
where the labeling $\Lambda_0$ and $\Lambda_2$ is just to denote that the cutoffs are not necessarily the same in each order.  $\bar{V}^{(^3P_0)}_{2\pi}$ is the ``simplified" version of the two-pion-exchange potential (TPE) in the $^3$P$_0$ channel defined in Ref.~\cite{Gasparyan:2022isg}, i.e., it comes from the partial-wave decomposition of a less divergent TPE,
\begin{equation}
\bar{V}_{2\pi}(\vec{p}',\vec{p})=V_{2\pi}(\vec{p}',\vec{p})\frac{(3m_{\pi})^2}{(3m_{\pi})^2+q^2},
\label{toy_2pe}
\end{equation}
where $m_{\pi}$ is the pion mass, $V_{2\pi}$ is the long-range part of TPE, and $q^2=(\vec{p}'-\vec{p})^2$.
$\bar{V}_{2\pi}$ has the same singularity as OPE and is accompanied by only one counter term with LEC $C^{NLO}_0$ at NLO. 
Note that $f_R$ in Eqs.~(\ref{toy1}) and (\ref{toy2}) denote any type of function that regulates the high-momentum part of the interaction. The following choice,
\begin{equation}
f_R(\Lambda)\equiv 0^n F(\Lambda)+(1-0^n) F(\Lambda/2), \label{toy3}
\end{equation}
where $n=0(1)$ for LO(NLO), coincides to the toy model illustrated in Fig.~2 of Ref.~\cite{Gasparyan:2022isg}. $F(\Lambda)$ denotes a sharp cutoff, i.e., 
\begin{equation}
F(\Lambda;p',p)=\theta(\Lambda-p') \theta(\Lambda-p). \label{toy4}
\end{equation}
Eq.~(\ref{toy3}) imposes a cutoff at NLO which is half of the LO value. One could nevertheless absorb this effect into the definition of $V_{NLO}$. The task now is to examine whether the PC based on DWBA is appropriate in this scenario. A straightforward check, as performed in Fig.~2 of Ref.~\cite{Gasparyan:2022isg}, shows a problematic RG behavior. However, as stated above, one needs to incorporate the corresponding uncertainty into account. In this toy model, the only way to cure the RG problem is to reverse the cutoff from $\Lambda/2 \rightarrow \approx\Lambda$ at NLO. When translated into Eq.~(\ref{dr}), it means the following needs to be satisfied for the perturbative PC to make sense:
\begin{widetext}
\begin{equation}
|\langle\psi_{LO}|(V^{(^3P_0)}_{2\pi}+C_0^{NLO}p'p)\left(F(\Lambda/2)-F(\Lambda)\right)|\psi_{LO}\rangle_i|\lesssim \mathcal{O}_n(M_{lo};\Lambda;M_{hi})\left(\frac{M_{lo}}{M_{hi}}\right)^{n+1}, \text{ for all } p_i\leq M_{hi},
\label{dr1}
\end{equation}
\end{widetext}
Here the value of $n+1$ depends on the power counting scheme, but in general $\gtrsim 1$, since it is at least one order suppressed than LO.
\begin{figure}[h]
\includegraphics[width=0.5\textwidth,clip=true]{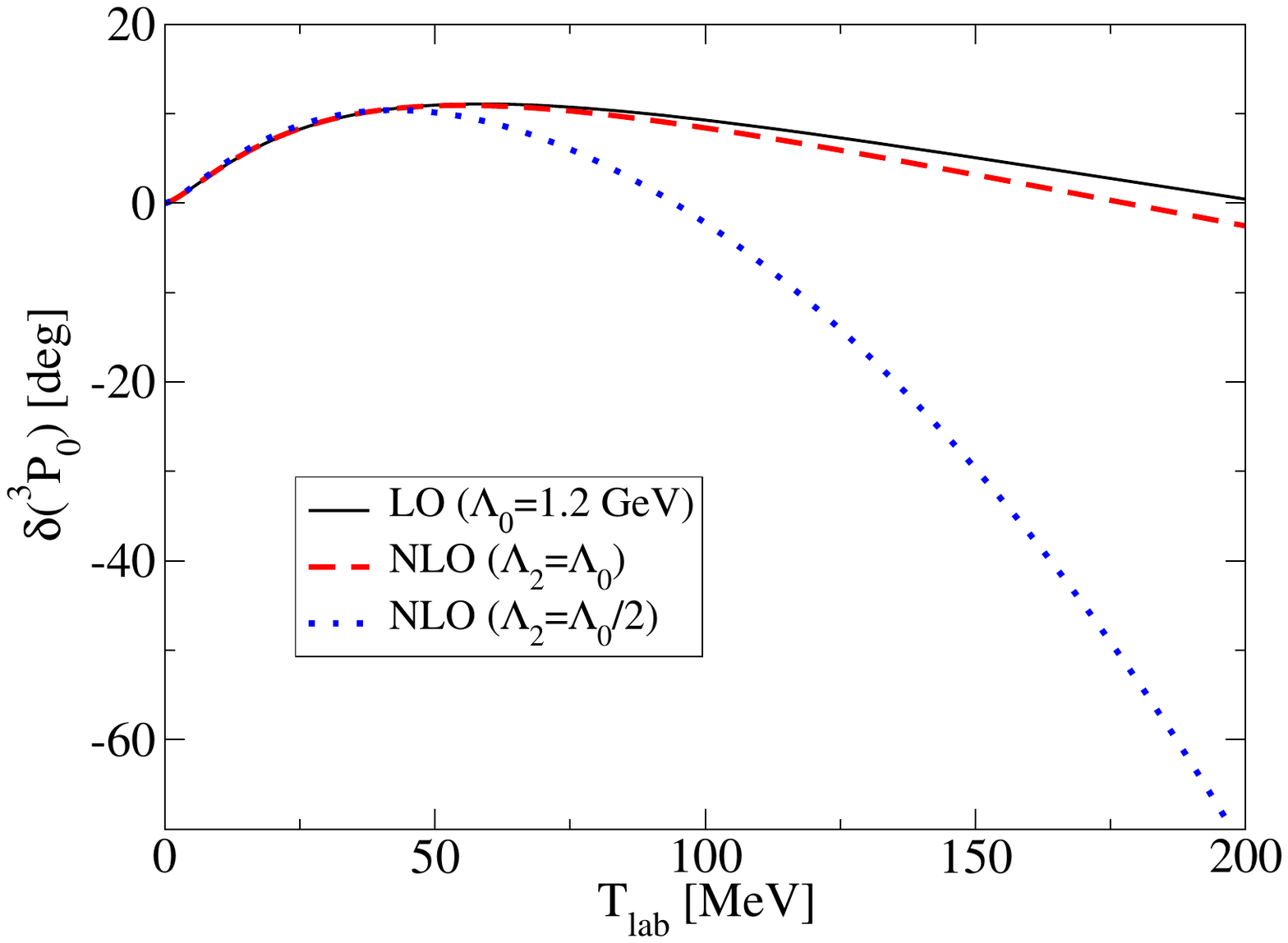}
\caption{Phase shifts at LO and NLO as a function of laboratory energy $T_{lab}$ for the toy model of Ref.~\cite{Gasparyan:2022isg}. The renormalization is performed at $T_{lab}=40$ MeV for all cases. Sharp cutoffs $\Lambda=\Lambda_0$ or $\Lambda_0/2$ are adopted.  }
\label{fig3}
\end{figure}

\begin{figure}[h]
\includegraphics[width=0.5\textwidth,clip=true]{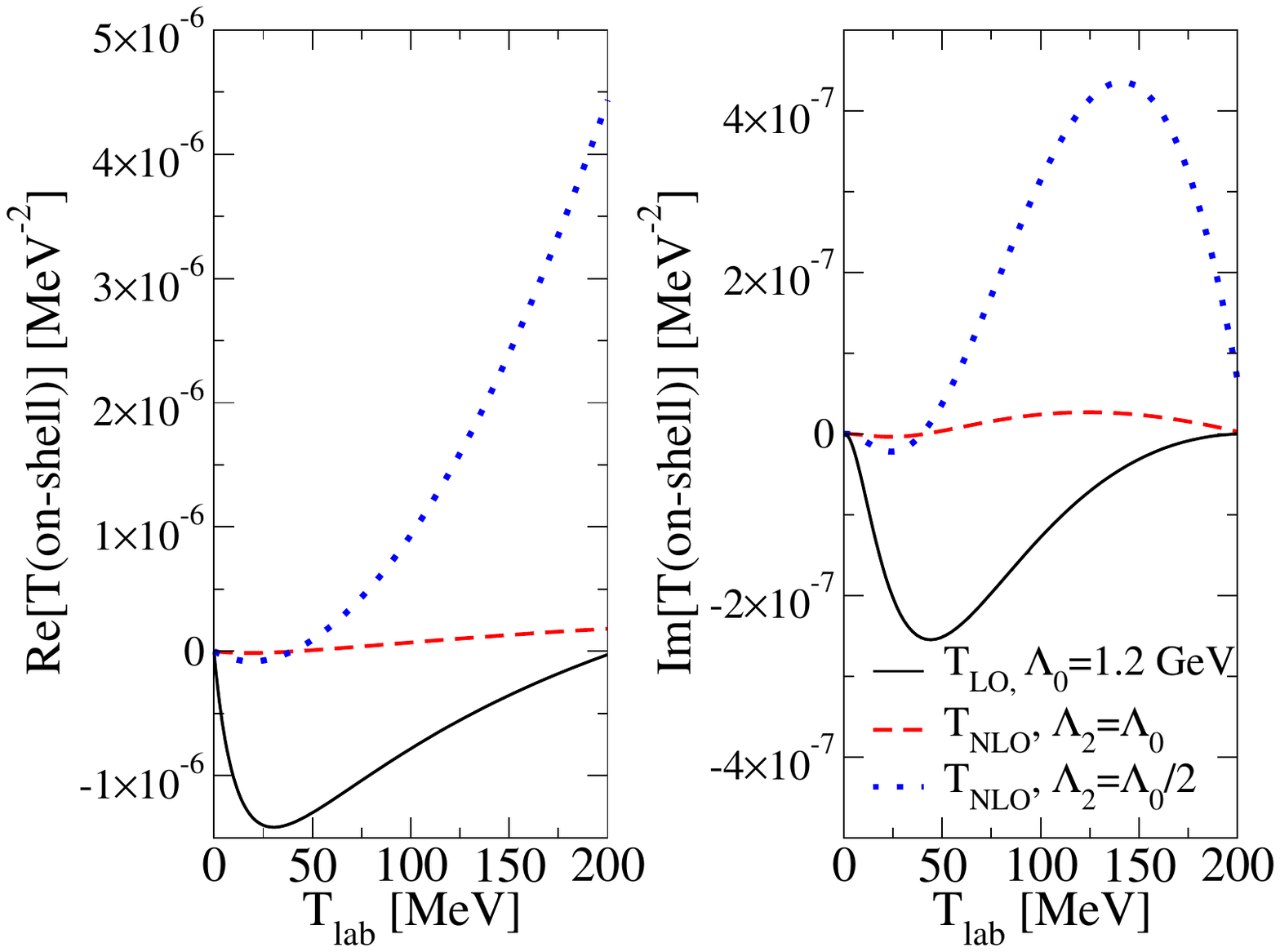}
\caption{The corresponding real and imaginary part of the on-shell T-matrix for the toy model where the phase shifts are presented in Fig.~\ref{fig3}.  }
\label{fig4}
\end{figure}
The phase shifts up to NLO are plotted as a function of $T_{lab}$ in Fig.~\ref{fig3}. As one can see, the adoption of $\Lambda_2=\Lambda_0/2$ at NLO leads to a more than $200\%$ difference for phase shifts at $T_{lab}>100$ MeV. As shown in Fig.~\ref{fig4}, the corresponding NLO T-matrix under $\Lambda_2=\Lambda_0/2$ becomes larger than the LO value, in contrast to the case where $\Lambda_2=\Lambda_0$ is adopted. Thus, the problematic pattern as shown in Fig.~2 of Ref.~\cite{Gasparyan:2022isg} will persist unless one prescribes that the NLO uncertainty is more than twice of the value at LO.
Therefore, one has to conclude that the DWBA-based PC indeed does not work for the toy model potentials listed in Eqs.~(\ref{toy1})-(\ref{toy2}).

\section{The PC proposed by Long and Yang in the $^3$P$_0$ channel}
\label{3p0}
In this section the same analysis is applied to the PC of Long and Yang in the $^3$P$_0$ channel~\cite{BY}, where the LO potential is the same as Eq.~(\ref{toy1}). At $O(Q^2)$, instead of Eq.~(\ref{toy2}), one has
\begin{equation}
V^{(^3P_0)}_{NLO}=[V^{(^3P_0)}_{2\pi}+C^{NLO}_0pp'+D_0^{NLO}pp'(p^2+p'^2)]f_R(\Lambda_0).
\label{nlo2}
\end{equation}
Note that the label ``NLO" here is adopted for convenience and is interchangeable with $O(Q^2)$ defined in Ref.~\cite{BY}. It is the first non-vanishing correction to LO in this channel, but not the first overall correction if one considers all partial-waves. Therefore, it is labeled as ``NNLO" in the PC of Long and Yang. 

As explained in Sect.~\ref{sec1}, the fact that the aforementioned RG issue~\cite{Gasparyan:2022isg} occurs only at an extremely narrow cutoff window can be an indication that it is merely an artifact of demanding unreasonable precision in the $O(Q^2)$ renormalization procedure. 
From Fig.~5 of Ref.~\cite{Gasparyan:2022isg}, it might already be clear that a direct combination of two sharp cutoffs, that is, with $f_b= f_a(\Lambda-0.001$ GeV) in Eq.~\ref{r} would solve the issue of exceptional cuts. In the following, I demonstrate that instead of a direct shift in the cutoff, 
a small uncertainty can be encoded using the following generalized regulator:
\begin{equation}
\mathfrak{F}_R=xf_a+(1-x)f_b, \label{ra}
\end{equation}
where
\begin{equation} \label{rab}
\begin{split}
f_a(\Lambda,p',p)&=\theta(\Lambda-p')\theta(\Lambda-p),\\
f_b(\Lambda,p',p)&=SG(\Lambda,p')SG(\Lambda,p), \\
\end{split}
\end{equation}
with
\begin{equation} \label{rdetail}
\begin{split}
SG(\Lambda,p^{(\prime)})&=1\text{ \ \ \ \ \ \ \ \ \ \ \ \ \ \ \ \ \ \ if $p^{(\prime)} \leq \Lambda/1.1$,}  \\ 
&=\exp(\frac{p^{(\prime)20}}{(\Lambda/1.1)^{20}}) \text{    if $p^{(\prime)} > \Lambda/1.1$}.
\end{split}
\end{equation}
Thus, the original sharp cutoff is replaced by a combination of sharp cutoff and Super-Gaussian regulator when one takes the choice $x=0$. Here the order of the Super-Gaussian regulator is taken to be very high so that $f_b$ is only slightly different from $f_a$.

\begin{figure}[h]
\includegraphics[width=0.5\textwidth,clip=true]{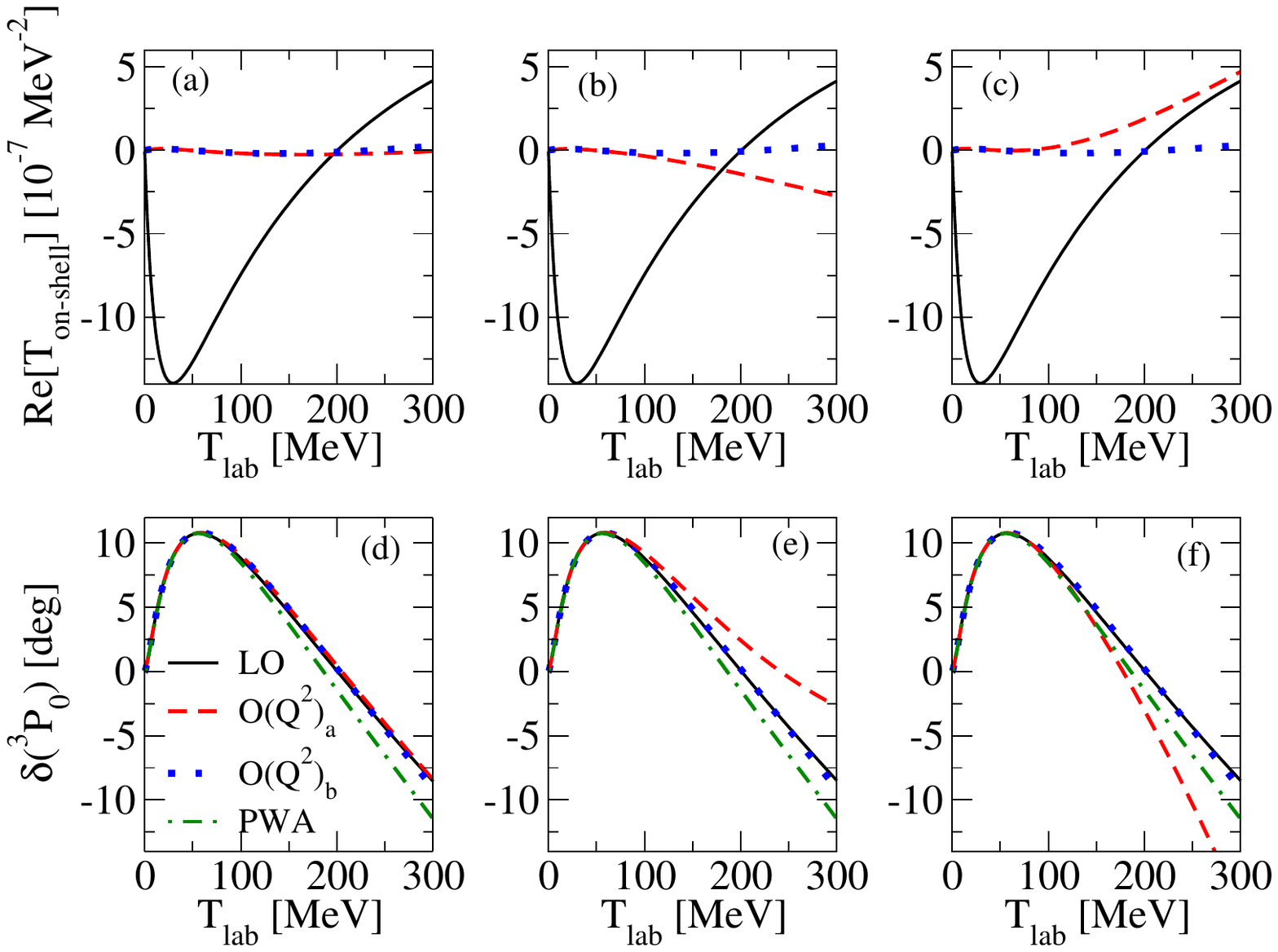}
\caption{Lower (upper) panels: Phase shifts (real part of the on-shell T-matrix) up to $O(Q^2)$ as a function of laboratory energy $T_{lab}$ for the Long and Yang PC~\cite{BY}. The renormalization is performed at $T_{lab}=50$ MeV at LO and $T_{lab}=25$, 50 MeV at $O(Q^2)$ for all cases. The left panel ((a) \& (d)) shows the result for a typical cutoff ($\Lambda=2640$ MeV), whereas the middle ((b) \& (e)) and right ((c) \& (f)) panels correspond to the cutoff values slightly below ($\Lambda=2698.08$ MeV) and above ($\Lambda=2699.18$ MeV) the exceptional point. $O(Q^2)_a$ and $O(Q^2)_b$ denote the regulators $f_a$ and $f_b$ in Eq.~(\ref{rab}). The LO regulator is $f_a$. }
\label{diff}
\end{figure}

In Fig.~\ref{diff}, the T-matrix and phase shifts at a typical cutoff $\Lambda=2.640$ GeV are plotted in the left column, while the same quantities near the problematic cutoff (where ``exceptional zero" occurs) are shown in the middle ($\Lambda=2.69808$ GeV) and the right ($\Lambda=2.69918$ GeV) column. Note that the exceptional cutoff $\Lambda\approx 2699$ MeV is very close to the cutoff where the LO LEC diverges (as shown in Fig.~\ref{fig2}). Nevertheless, the LO amplitude is well-defined and can be obtained numerically by the subtractive renormalization scheme~\cite{Yang:2009kx}. Here I adopted the same renormalization procedure as performed in Ref.~\cite{Gasparyan:2022isg}, i.e., the LECs are fixed at $T_{lab}=50$ MeV at LO and $T_{lab}=25, 50$ MeV at $O(Q^2)$. 

In Fig.~\ref{diff}, the issue of ``exceptional zero" is reflected in the red dashed-line in the middle and right columns, where the phase shifts diverge from their ``typical cutoff values" (the red dashed-line in (d)). This is because extreme LECs are required to generate finite corrections as the matrix elements go to zero. On the other hand, the problem disappears when the value $x=0$ is taken in Eq.~(\ref{ra}), which is demonstrated by the fact that the blue dotted line remains unchanged throughout all panels in Fig.~\ref{diff}. 
Furthermore, panel (a) shows that, under a typical cutoff, the maximum uncertainty generated by Eq.~(\ref{ra})---which corresponds to a direct change from adopting a regulator that is fully $f_a$ to $f_b$---produces a difference that is negligible compared to the LO amplitude and is certainly smaller than the uncertainty prescribed by the PC at $O(Q^2)$. In other words, the variation of $x \in [0,1]$ indeed corresponds to taking into account an uncertainty allowed by EFT.

Note that as long as $x$ has a fixed value throughout all $\Lambda$, ``exceptional cutoffs" will only be shifted but still occur. 
Thus, $x$ should be varied whenever necessary to avoid the problem. 
One could doubt that adjusting $x$ within a limited range is a choice depending on the outcome, but it is a justified choice due to the nature of DWBA. 
In fact, once Eq.~(\ref{dr}) is satisfied, not fixing $x$ to an exact value corresponding to not taking the perturbative NLO correction too exact---as demonstrated in Eq.~(\ref{eq1}).

\begin{figure}[h]
\includegraphics[width=0.5\textwidth,clip=true]{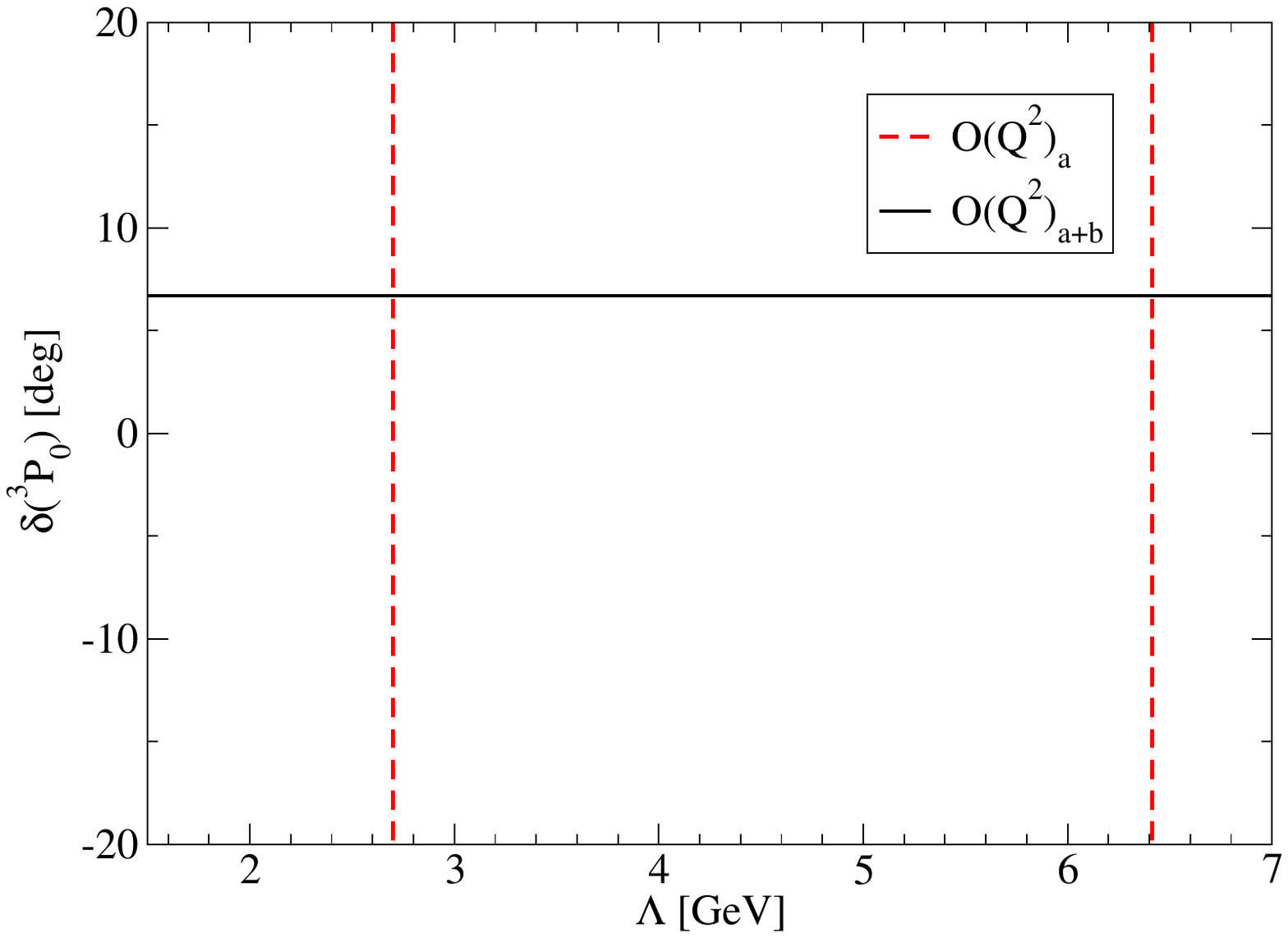}
\caption{Cutoff dependence of $^3$P$_0$ phase shift at $T_{lab}=130$ MeV calculated using Long and Yang PC at $O(Q^2)$. The red-dashed lines denote the diverged results at the exceptional cutoffs using the regulator $f_a$ in Eq.~(\ref{rab}). The black solid line denote the results from the generalized regulator with an adjustable $0\leq x\leq 1$ whenever necessary to avoid fine-tune related problems, as defined in Eq.~(\ref{ra}).  }
\label{diff2}
\end{figure}

By coupling this uncertainty into the renormalization procedure, 
one is able to retain a converged pattern in both T-matrix\footnote{Note that the LO to NLO pattern of the imaginary part of T-matrix (which was not plotted in Fig.~\ref{diff}) presents the same feature as their real part.} and phase shifts, therefore avoiding the appearance of extreme LECs.
The phase shift as a function of $\Lambda$ is plotted in Fig.~\ref{diff2}, where $x$ is varied between $0\sim1$ whenever necessary to avoid taking an accuracy that exceeds the prescribed PC.
As one can see, the issue of ``exceptional zero" in the Long and Yang PC represented by the two vertical red-dashed lines disappears when one avoids taking an unreasonable accuracy beyond $O(Q^2)$ in the renormalization process.

\section{Application on the conventional WPP}
\label{wpc}
One might wonder whether the above ingredient, i.e., by taking uncertainty into account when performing renormalization, can resolve the RG issues of WPP. At LO, WPP has been shown to miss appropriate counter terms in those singular and attractive channels~\cite{Nogga:2005hy}.
The problem is of mathematical origin, i.e., singular potentials render the solution of Schr\"odinger equation undetermined without the presence of a proper boundary condition. Note that the problematic cutoff (where the phase shift deviates from the typical value) first appears at $\approx 1000$ MeV and spans at least for 200 MeV, as demonstrated in the $^3$P$_0$ channel panel of Fig.~9 in Ref.~\cite{Nogga:2005hy} and Fig.~\ref{wpclo} below.  

Starting at NLO under WPP, the problem persists in two types of pathology. The first has the same origin, i.e., the boundary condition, as LO. For example, the TPE($Q^2$) in $^3$P$_0$ channel is repulsive at the origin and dominates OPE at short distances. Thus, imposing a counter term ``$C^{NLO}p'p$" to renormalize OPE+TPE($Q^2$) causes an RG problem, as shown in Fig.~\ref{wpcnlo} below. The second type of RG problem concerns a Wigner-bound-like effect, which occurs when two counter terms are included non-perturbatively. As the increase of $\Lambda$, the short-range counter terms dominate over pion-exchange and one approaches the regime governed by Wigner-bound~\cite{wigner,wigner2}, which is reflected in, e.g., the NLO S-waves under WPP~\cite{Yang:2009pn}. Note that the problematic cutoff starts at $\Lambda\approx 1000$ MeV and spans at least for 100 MeV for both NLO($Q^2$) and NNLO($Q^3$).

The broadness of the problematic cutoff ($\sim 100-200$ MeV for $\Lambda\approx 1000$ MeV) under WPP is in sharp contrast to those found in the Long and Yang PC ($\sim 0.1$ MeV for $\Lambda\approx 1000$ MeV) without considering the uncertainty.
As demonstrated in the previous sections, the uncertainty allowed by EFT at a given order may be encoded into a small change in the regulator or cutoff. It is of interest to see whether those problematic cutoffs under WPP can be eliminated by the uncertainty argument. 
A naive estimate can be performed as follows. Denoting $\Lambda$ and $\Delta$ the starting value and the broadness of the problematic cutoff, and $Q^d$ the leading divergence of the potential in momentum space. Then in the spirit of Eq.~(\ref{dr}), the minimum requirement for legally including a small uncertainty in terms of the cutoff/regulator to eliminate the problematic region is 
\begin{equation}
\frac{(\Lambda+\Delta)^d-\Lambda^d}{\Lambda^d}\leq(\frac{M_{lo}}{M_{hi}})^{n+1},
\label{dr3}
\end{equation}
where $n$ is the order specified by the PC. Note that the above counts only the difference of the regulator's impact on the potential at the shortest range. The actual requirement involves an integral over $\int_{\Lambda}^{\Lambda+\Delta} dp$ for the potential together with the wavefunctions on the left-hand side, which generally results in a much stricter requirement. Thus, the left-hand side might need to be much smaller than the right-hand side. 
Applying the above to WPP at NLO($Q^2$), one has
\begin{equation}
\frac{(\Lambda+\Delta)^2-\Lambda^2}{\Lambda^2}=2\frac{\Delta}{\Lambda}+\frac{\Delta^2}{\Lambda^2}\approx 2\frac{\Delta}{\Lambda},
\label{dr4}
\end{equation}
where the last step assumes $\Delta<<\Lambda$. Plugging $\Lambda\approx 1000$ MeV and $\Delta\approx 100$ MeV, the left-hand side of Eq.~(\ref{dr3}) $\approx 0.2$. Meanwhile, the right-hand side of Eq.~(\ref{dr3}) gives $(1/3)^2=1/9\approx 0.11$. Applying the same $\Lambda$ and $\Delta$ to NNLO($Q^3$) results in a worse violation of Eq.~(\ref{dr3}). Since the minimum requirement is not even met, this suggests that the problem of WPP cannot be resolved by taking uncertainty into account.  

The above conclusion is further verified via actual calculations of the $^3$P$_0$ channel NN scattering under WPP, where the problematic phase shifts at LO and up to NLO($Q^2$) are presented in Fig.~\ref{wpclo} and Fig.~\ref{wpcnlo}, respectively. In this channel, WPP does not prescribe a counter term in LO. At NLO($Q^2$), a counter term in the form of $C_2p'p$ is presented, which is fitted to reproduce Nijmegen PWA at $T_{lab}=40$ MeV. As shown in Eq.~(\ref{dr}), to attribute these undesirable phase shifts between $\Lambda=900-1000$ MeV to an unreasonable precision demand, one would need to show that the effect of shifting $\Lambda\approx950\pm 50$ MeV is smaller than the theoretical uncertainty. Meanwhile, as shown in Fig.~\ref{wpc_t}, such changes of $\Lambda$ generate differences in amplitude (denoted as the blue dotted-lines) with values comparable to typical size of the amplitude at that order. 
Taking 1/3 as the typical expansion parameter of chiral EFT and the fact that WPP prescribes a suppression O$(Q^2)$ between the NLO and the LO amplitude, the blue dotted-line in the lower panel of Fig.~\ref{wpc_t}---which presumably should be an NLO effect---not only exceeds 1/9 of the LO value, but is larger than 1/3 around $T_{lab}=100$ and after $T_{lab}\geq270$ MeV. This violation of the prescribed PC clearly exceeds a reasonable tolerance governed by ``weak naturalness"~\cite{Wesolowski2016,Melendez:2020ikd}. 
Therefore, WPP remains non-renormalizable even with uncertainty taken into account.      

\begin{figure}[h]
\includegraphics[width=0.5\textwidth,clip=true]{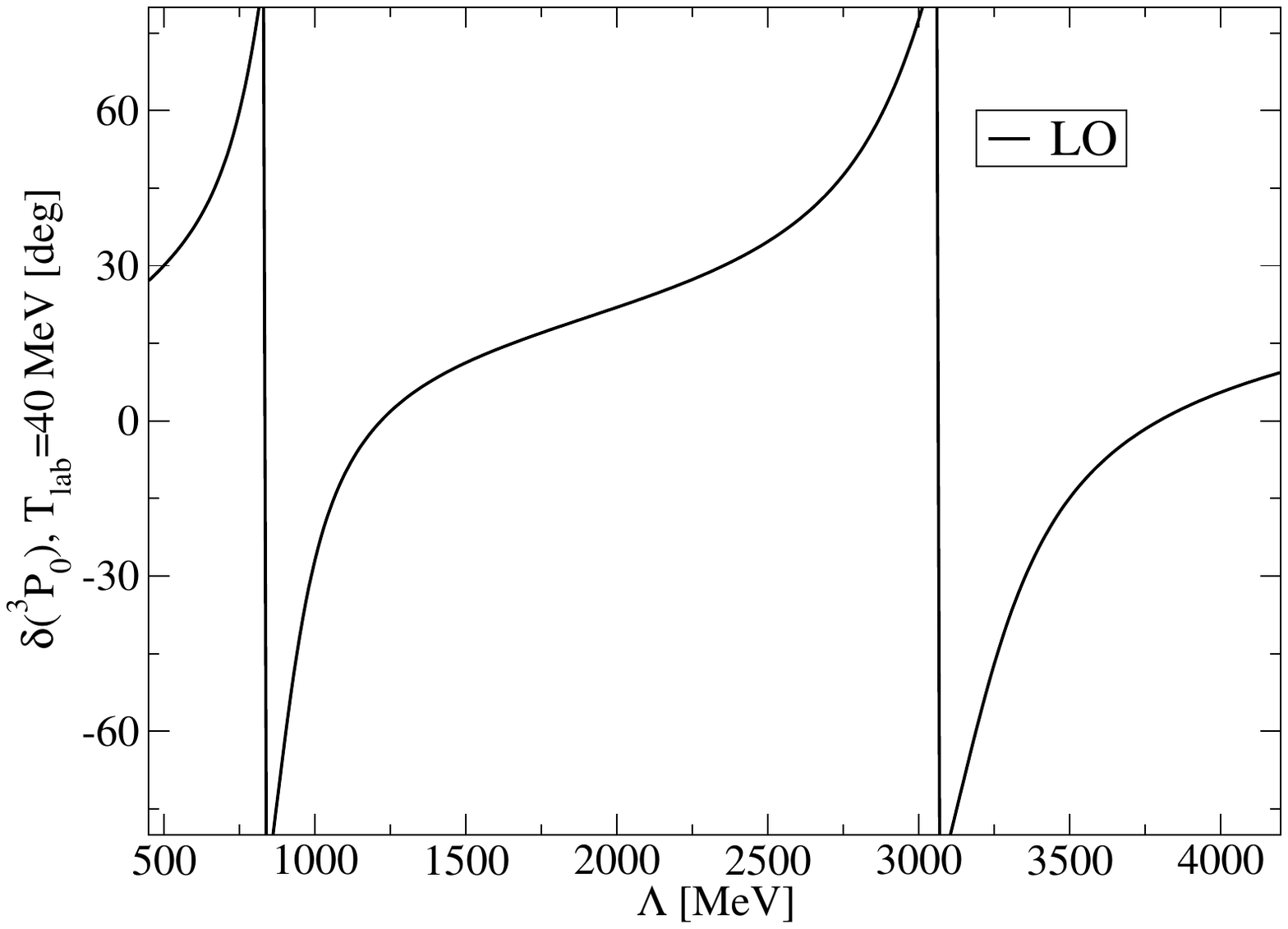}
\caption{Cutoff dependence of $^3$P$_0$ phase shift at $T_{lab}=40$ MeV calculated using WPP at LO. The potential is OPE with a regulator $f_R=\exp{[-\frac{(p'^4+p^4)}{\Lambda^4}]}$ and without counter terms.  }
\label{wpclo}
\end{figure}
\begin{figure}[h]
\includegraphics[width=0.5\textwidth,clip=true]{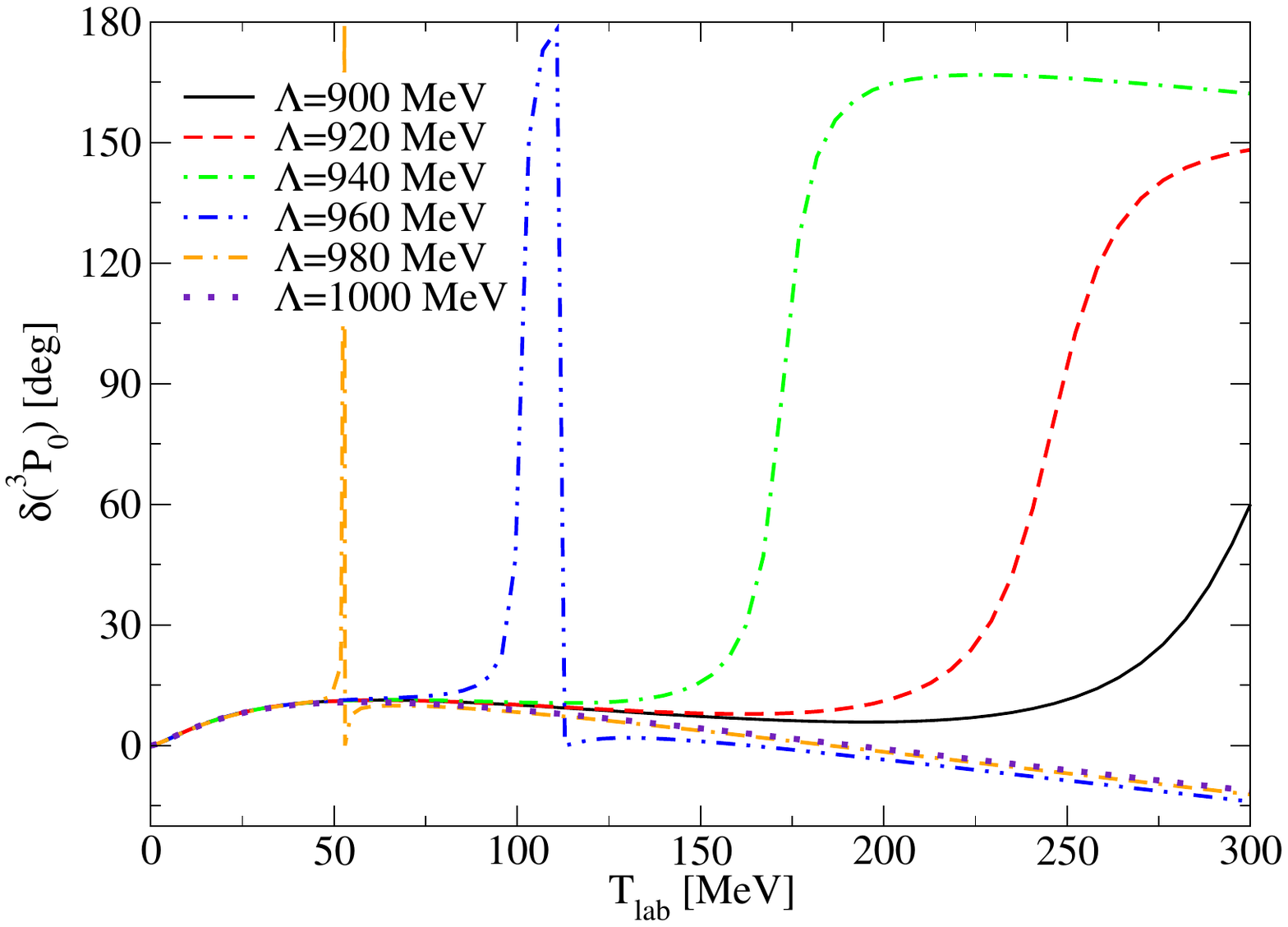}
\caption{$^3$P$_0$ phase shift as a function of laboratory energy $T_{lab}$ calculated using WPP at NLO$(Q^2)$. The potential is OPE+TPE($Q^2$)+$C_2p'p$, with a regulator $f_R=\exp{[-\frac{(p'^4+p^4)}{\Lambda^4}]}$ and $C_2$ renormalized to fit Nijmegen PWA at $T_{lab}=40$ MeV. All phase shifts $\delta>180$ [deg] are plotted as $\delta-180$ [deg]. }
\label{wpcnlo}
\end{figure}

\begin{figure}[h]
\includegraphics[width=0.5\textwidth,clip=true]{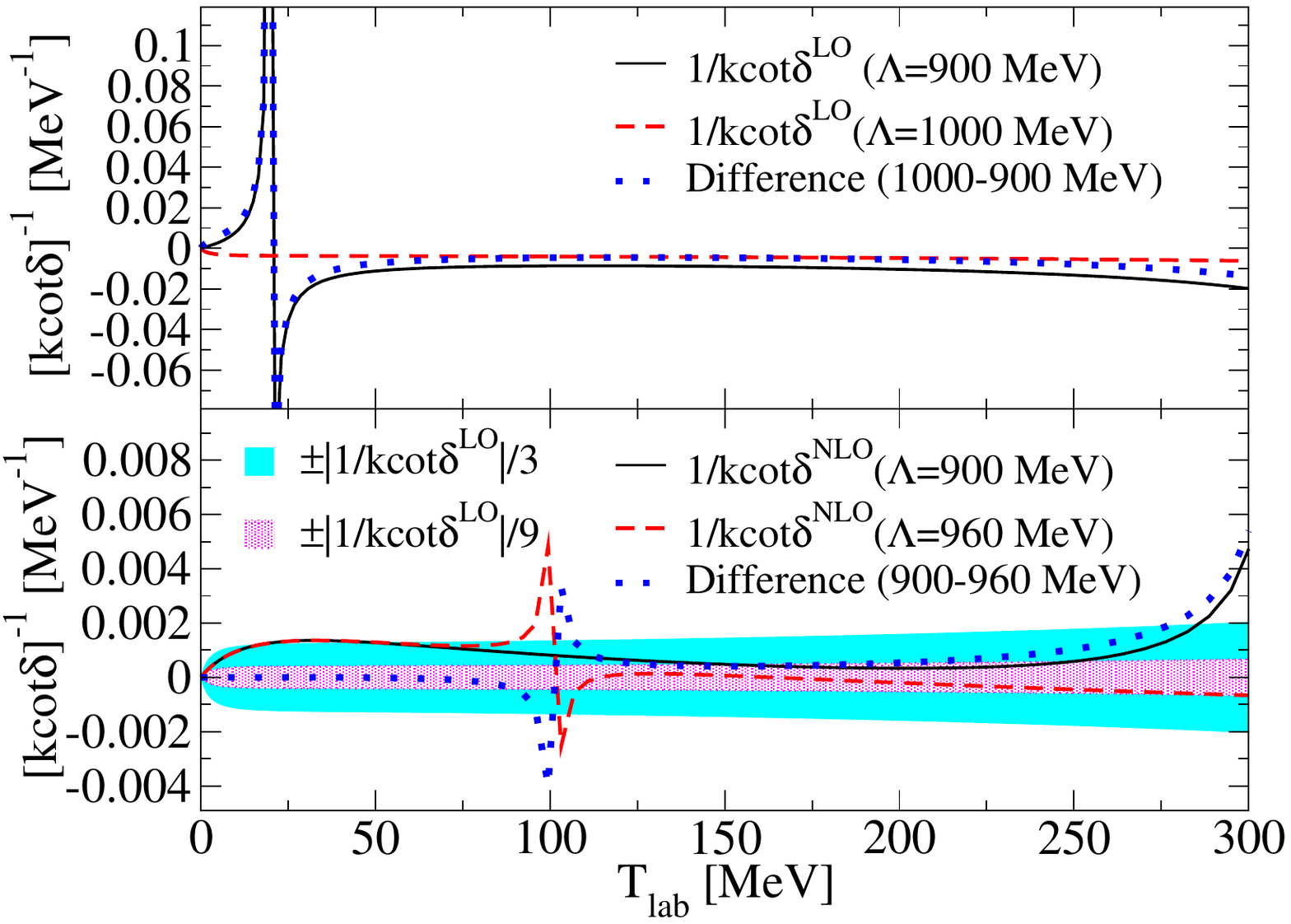}
\caption{Upper (lower) panels: Real part of the $[k\cot{\delta}]^{-1}$ at LO (up to NLO$(Q^2)$) as a function of laboratory energy $T_{lab}$ obtained by WPP. The renormalization at NLO is performed at $T_{lab}=40$ MeV at NLO. The blue dotted-lines are the difference due to the two different cutoffs. The cyan and magenta regions represent $\frac{1}{3}$ and $(\frac{1}{3})^2$ of the typical LO value (obtained with $\Lambda=1000$ MeV), respectively. }
\label{wpc_t}
\end{figure}

\section{Further justifications and the role of regulators}
\label{regu}

The main point of this work can be summarized as follows:

``To avoid an exceptional zero, the subleading amplitudes---especially those obtained through DWBA---are subjected to modifications up to the magnitude allowed by the PC."

Note that such an ad hoc modification does not violate any EFT principle. Essentially, the construction of an EFT generally involves a trial-and-error process. With the origin of the exception zero analyzed in detail, one could anticipate the cutoffs where they occur by examining the subleading matrix elements as a function of momentum $p_0$. Then, necessary action is taken to prevent the aforementioned problem. In this way, no predictive power is lost.

It is also evident that the regulator proposed in Eq.~(\ref{r}) is merely one possibility to solve the problem. There could exist other regulators which avoid the exceptional cutoff issue. For example, the issue might be avoided/mitigated by the regulators adopted in Refs.\cite{Valdper,Valdperb,manolo}. From a more general point of view, reshuffling the fitting as performed in Ref.~\cite{Peng:2024aiz} is equivalent to modifying the amplitudes via the regulator listed in Eq.~(\ref{r}). Thus, as long as the modification is small compared to the allowed uncertainty, one can satisfy all the EFT requirements.

\section{Summary}
\label{con}
A peculiar type of RG issue under DWBA as discovered in Ref.~\cite{Gasparyan:2022isg} is further studied by considering the theoretical uncertainty inherent in its NLO amplitude. It is shown that the problem concerning ``exceptional zero" can have two sources: (i) from enforcing an accuracy exceeding the prescribed PC in the wavefunctions/observables at the given order, or, (ii) from a wrong arrangement of PC. In general, the broadness of the problematic cutoff serves as a guide to distinguish the two cases. 
With a detailed investigation, it is shown that the RG issue for PC of Long and Yang belongs to the former, while the toy model in Ref.~\cite{Gasparyan:2022isg} belongs to the latter.
By applying the same argument to WPP, it is found that the RG/Wigner-bound problem cannot be resolved by taking uncertainty into account.

\vspace{0.5cm}
\begin{acknowledgments}
I thank Bingwei Long, M. Pavon Valderrama and U. van Kolck for
useful discussions and suggestions. This work was supported by the
the Extreme Light Infrastructure Nuclear Physics (ELI-NP) Phase II, a project co-financed by the Romanian Government and the European Union through the European Regional Development Fund - the Competitiveness Operational Programme (1/07.07.2016, COP, ID 1334);  the Romanian Ministry of Research and Innovation: PN23210105 (Phase 2, the Program Nucleu); and ELI-RO\_RDI\_2024\_AMAP, ELI-RO\_RDI\_2024\_LaLuThe, ELI-RO\_RDI\_2024\_SPARC of the Romanian Government; and the European Union, the Romanian Government and the Health Program, within the project "Medical applications of high-power lasers - Dr. LASER"; SMIS Code: 326475.
 I acknowledge PRACE for awarding us access to Karolina at IT4Innovations, Czechia under project number EHPC-BEN-2023B05-023 (DD-23-83); IT4Innovations at Czech National Supercomputing Center under project number OPEN-34-63 and OPEN24-21 1892; Ministry of Education, Youth and Sports of the Czech Republic through the e-INFRA CZ (ID:90140) and CINECA under PRACE EHPC-BEN-2023B05-023.


\end{acknowledgments}

\bibliography{res_epel} 
\bibliographystyle{apsrev4-1}

\end{document}